\documentclass[doublecol]{epl2}

\usepackage{indentfirst}
\usepackage[english]{babel}
\usepackage{graphicx}
\usepackage{latexsym,amsmath,amssymb,amsfonts,amsthm,enumerate}

\def\bef{\beta_{\rm eff}}

\newcommand{\f}{\footnote}
\newcommand{\nn}{\nonumber}

\renewcommand{\(}{\left (}
\renewcommand{\)}{\right )}
\newcommand{\la}{\langle}
\newcommand{\ra}{\rangle}

\newcommand{\eq}[2][]{
\begin{equation}
#2 \label{#1}
\end{equation} }

\newcommand{\ee}[2][]{
\begin{align}
#2 \label{#1}
\end{align}}

\newcommand{\beff}{\beta_{\text{eff}}}
\newcommand{\bpq}{\beff^{*}}

\title{Quantum Quench from a Thermal Initial State.}

\author{Spyros Sotiriadis\inst{1} \and Pasquale Calabrese\inst{2} \and John Cardy\inst{1,3}}
\shortauthor{S.~Sotiriadis, P.~Calabrese, J.~Cardy}

\institute{
  \inst{1} Oxford University, Rudolf Peierls Centre for Theoretical Physics, 1 Keble Road, Oxford OX1 3NP, UK \\
  \inst{2} Dipartimento di Fisica Enrico Fermi, Universit\`a di Pisa, Largo Bruno Pontecorvo 3, 56127 Pisa, Italy \\
  \inst{3} All Souls College, High Street, Oxford OX1 4AL, UK
}


\abstract{We consider a quantum quench in a system of free bosons,
starting from a thermal initial state. As in the case where the
system is initially in the ground state, any finite subsystem
eventually reaches a stationary thermal state with a
momentum-dependent effective temperature. 
 {We
find that this can, in some cases, even be lower than the initial
temperature.} We also study lattice effects and discuss more general types of quenches.}

\begin{document}

\maketitle

\section{Introduction}

The problem of a \emph{quantum quench}, i.e. an instantaneous change in the
parameters that determine the dynamics of a quantum system, has been recently
studied in various theoretical models.
These include systems of free bosons or fermions in which the
quenched parameter is the energy gap \cite{1,2,3},
integrable models described by conformal field theory
\cite{1,2,4,5,6} and exactly solvable spin
chains  {\cite{4,7,8}}.
Several studies have also been carried out in non-integrable
models using numerical methods or approximation schemes
 {\cite{9,10,11,11b 
}}.
Perhaps one of the most interesting results
is that in the \emph{thermodynamic limit} 
it is possible under quite general conditions that for large times an
arbitrarily large \emph{subsystem} tends to a stationary state with thermal
characteristics, a process we can call \emph{thermalization}
 {\cite{1,2,3,12}}. \looseness=-1

To be more specific, let us imagine a system of coupled harmonic oscillators
or equivalently a \emph{free} field theory, described by a general dispersion
relation with some energy gap or ``mass" $m_0$ and maximum group velocity of
excitations $c$. Assume that the system initially lies on the ground state
of the initial hamiltonian $H_0$ and at time $t=0$ the mass is suddenly
changed from $m_0$ to a different 
value $m$. For $t>0$, the state of the system evolves according to
quantum mechanics (\emph{unitary} evolution), i.e.~it is fully
isolated from its environment. After this quench there is an
extensive energy excess in comparison with the ground state of the
final hamiltonian $H$, which is distributed over the excitation
levels of $H$. It then turns out that the two-point correlation
function, which in free systems contains all the information
required to determine their state, acquires for sufficiently large
times the form of the correlation function of a system at
\emph{thermal} equilibrium with a \emph{momentum-dependent}
effective temperature $\bef (k)$ \cite{1,2,13}. Note
that this momentum dependence is expected for a free theory since
each momentum mode-excitation evolves independently from the
others and can therefore thermalize to a different temperature.
For any local observable, the quantum interference between all the
momentum modes gives rise to a single effective temperature,
which, however, may be dependent on the chosen observable.
However, the large distance behaviour of the two-point correlation
function is determined only by the zero-momentum mode $\bef (0)$.
In the so called \emph{deep} quench limit where $m_{0} \gg m$, the effective
temperature $\bef^{-1}(k)$ is asymptotically equal to $m_0/4$ and becomes
momentum independent \cite{1,2}. \looseness=-1

In most studies made so far,
the initial state of the system is the ground state of $H_{0}$.
Here instead, we assume that, before the quench, the system is
prepared in a \emph{thermal} state. Then we perform the quench
exactly as before keeping the system isolated from the environment
(the coupling to the environment is studied in \cite{14}). For
brevity we can call this a \emph{thermal quantum quench}, in
contrast to the previous case that we will call a \emph{pure
quantum quench}.  {Few other studies 
including the effect of a thermal initial state exist \cite{8,11b,13b}, 
but consider different kinds of quenches. }
The solution of this problem is in general
prohibitively difficult and effective numerical methods must be
used to extract useful physical quantities. However, in free field
theory the solution is particularly simple and allows us to
investigate the effect of the initial temperature $\beta_{0}$ on
the time evolution of the system. Especially we can study the
crossover from the ``\emph{hot}'' limit, i.e. when the initial
temperature dominates, to the ``\emph{cold}'' limit when it can be
ignored, in which case we recover the pure quantum quench results.  \looseness=-1

It turns out that after such a thermal quench the system also
thermalizes in the sense described earlier. An intuitive guess
would perhaps be that the final effective temperature $\bef^{-1}$
is just the sum of the initial temperature $\beta_{0}^{-1}$ and
that corresponding to an equal \emph{pure} quench of the mass
${\bef^{*}}^{-1}$. However this is not true  {and $\bef^{-1}$ can instead be lower} than $\beta_{0}^{-1}$.
In the hot deep quench limit ($\beta_0^{-1} \gg m_{0} \gg m$), for
example, \emph{the final temperature is shown to be half of the
initial one}.

In the present paper we provide the solution of the problem of a
thermal quantum quench in a system of free bosons. 
We start with a simple harmonic
oscillator where we calculate the analogue of the correlation
function as a function of time. Then we use this to find the
correlation function in a bosonic free field theory  {with 
a relativistic dispersion relation} and study its
large time as well as real space behaviour. We thus identify its
stationary part and, by comparison to the thermal expression,
derive the effective temperature, whose properties in several
limits are discussed.  {Furthermore we 
investigate} several other types of quenches and derive general
expressions for the momentum distribution of excitations in
lattice models also extending to the fermionic case. 

\section{Harmonic oscillator}

We first consider a single harmonic oscillator that initially lies
in a thermal state with temperature $\beta_0$. Then at $t=0$ we
quench its frequency from $\omega_0$ to $\omega$.
The time evolution of the position operator after the quench is
given by the Heisenberg equations of motion $\ddot{x}(t) +
\omega^2 x(t) = 0$ with solution $x(t) = x(0) \cos \omega t + p(0)
{\sin \omega t}/{\omega}$. The computation of the time ordered
correlator $\mathcal{T} \{x(t_1) x(t_2)\}$ thus reduces to finding
the values of $x^2$, $xp$, $px$ and $p^2$ in the initial state.
These values are thermal statistical averages of quantum
expectation values given in general by $\mathcal{{O}}(\beta) =
\text{Tr}(e^{-\beta H}\mathcal{\hat{O}})/Z(\beta) =
\sum_n{e^{-\beta E_n}\langle n|\mathcal{\hat{O}}|n \rangle
}/Z(\beta)$.
For a simple harmonic oscillator of frequency $\omega_0$ and at
temperature $\beta_0$, since the energy levels are $E_n =
(n+\tfrac{1}{2})\omega_0$, the partition function is $Z(\beta_0) =
1/(2 \sinh{{\beta_0} {\omega_0}/2})$. From the equipartition
theorem and the equation $E(\beta) = - {\partial} (\ln
Z({\beta})) / {\partial {\beta}}$ for the energy of a thermal
state, or by direct calculation, we find 
\eq{ \langle x^2
\rangle_{\beta_0} = \frac{1}{2 \omega_0} \coth \frac{\beta_0
\omega_0}{2} \; ,\quad\langle p^2 \rangle_{\beta_0} =
\frac{\omega_0}{2} \coth \frac{\beta_0 \omega_0}{2}\,. \nn} Also
$[x,p] = i$ and $ \langle \{x,p\} \rangle_{\beta_0} = 0 $. Using
all of the above equations we can show that the time ordered
correlator is 
\ee[pro]{ C_{\beta_{0}}&(t_1,t_2) = \mathcal{T}
\{x(t_1) x(t_2)\} =
\frac{1}{2 \omega} e^{ -i \omega |t_1 - t_2|} + \nn \\
& + \left[ \frac{\omega_0}{4} \left(\frac{1}{\omega_0^2} +
    \frac{1}{\omega^2}\right) \coth \frac{\beta_0 \omega_0}{2}
-\frac{1}{2 \omega}\right] \cos \omega (t_1-t_2) + \nn \\
& + \frac{\omega_0}{4} \left(\frac{1}{\omega_0^2} - \frac{1}{\omega^2}\right)
\coth \frac{\beta_0 \omega_0}{2}  \cos \omega (t_1+t_2)\,.}
\looseness=-1  {The first term is the Feynman propagator in 0+1 
dimensions and is the only one that survives if 
$\omega_{0} = \omega$ and $\beta_{0}\to \infty$.}
The breaking of time-translational invariance due to the quench is apparent in
the last term that 
depends on $t_{1}+t_{2}$. Also notice that for $\beta_{0} \to
\infty$ we recover the pure quantum quench propagator, since the
initial state is then the ground state of the initial hamiltonian 
\ee[propq]{C(t_1,t_2) = & \frac{1}{2 \omega} e^{ -i \omega |t_1 - t_2|} 
+ \frac{(\omega - \omega_{0})^{2}}{4 \omega_{0} \omega^{2}} \cos
\omega (t_1-t_2) +  \nn \\
&+ \frac{\omega^{2} - \omega^{2}_{0}}{4 \omega_{0}
\omega^{2}} \cos \omega (t_1+t_2)\,. } 
Obviously the harmonic
oscillator does not thermalize but oscillates. However, in a free
field theory, we shall argue that the interference between
momentum modes needed to calculate local observables is
responsible for effective thermalization.

From the above results we find that the energy of the oscillator
before the quench is \ee[E0]{\la H_{0} \ra_{\beta_{0}} =
\tfrac{1}{2} \omega_0 \coth (\beta_0 \omega_0/2) } while after
the quench it is \ee[E]{\la H \ra_{\beta_{0}} =
\frac{\omega^{2}+\omega_{0}^{2}}{4 \omega_0} \coth (\beta_0 \omega_0/2) } i.e. it changes by a factor $(E - E_{0})/E_{0} =(
(\omega / \omega_{0})^{2}~-~1)/2$. This means that the work done on the system at the quench is positive or negative, depending on whether the frequency increases or decreases respectively. Notice
that for $\omega\ll \omega_{0}$ the system loses half of its
energy.

\section{Bosonic free field theory}

Let us now move on to a bosonic free field theory with hamiltonian of the general form 
\ee[ham]{H = \int{d^{d}k \; \(\frac{1}{2} \pi_{k}^{2} + \frac{1}{2} \omega^{2}_{k} \phi_{k}^{2}\)},}
and assume a relativistic dispersion relation  {
$\omega_{k}^{2} =c^{2} k^{2}+m^{2}c^{4} $} 
in which we quench the mass from $m_{0}$ to $m$. We will later discuss other
types of quenches with different dispersion relations or different quench
parameters, but we now specialize to this case which
contains most of the physical features of the general cases.
For brevity we set the speed $c=1$. 

The propagator in the mixed momentum-time representation is just
that of a single harmonic oscillator (\ref{pro}) with frequencies
$\omega_{k}$ and $\omega_{0k}$  {defined by $m$ and $m_0$ respectively.} 
The real space propagator is the Fourier
transform of the latter and we can easily check, using the
stationary phase method, that for large times and finite
separations the integration over all momenta leads to the
$(t_{1}+t_{2})$-dependent part of the propagator decreasing with
time under quite general conditions (for $m \neq 0$ or even for
$m=0$ in 3$d$ \f{Even though this condition does not hold for
$m=0, d=1$, we can still talk about thermalization in $1d$
critical systems, since 
what is physically meaningful is not the propagator, but the correlation
function of \emph{vertex} operators, 
 {i.e. imaginary exponentials of the field operator}, and the latter does become stationary \cite{2}.}). This property is already known for the
pure quench propagator (\ref{propq}) \cite{1,2,13} and
we now simply observe that it also holds in the more general
thermal case (\ref{pro}). According to the above, the correlation
function tends for large times to the stationary form
\ee[pro-st]{&\tilde{C}_{\beta_{0}}(k;t_1,t_2) =
\frac{1}{2 \omega_{k}} e^{ -i \omega_{k} |t_1 - t_2|} + \nn \\
& + \left[ \frac{\omega_{0k}}{4} \left(\frac{1}{\omega_{0k}^2} +
\frac{1}{\omega_{k}^2}\right) \coth \frac{\beta_0 \omega_{0k}}{2}
-\frac{1}{2 \omega_{k}}\right] \cos \omega_{k} (t_1-t_2). }
Furthermore one can compare this to the thermal or Matsubara
propagator in real time which gives the correlations in a system
at thermal equilibrium \ee{G_{\beta}(k;t_1,t_2) =  \frac{1}{2
\omega_{k}} e^{-i\omega_{k} |t_{1}-t_{2}|} + \frac{\cos \omega_{k}
(t_1-t_2)}{\omega_{k} (e^{\beta \omega_{k}}-1)}.} Again as in the
pure quench case, the stationary propagator (\ref{pro-st}) is of
thermal form, where the effective temperature $\beff$ is
determined by equating the coefficients of $\cos \omega_k (t_1 -
t_2)$ in the compared expressions,
giving
\eq[cond]{
\bef(k)=\frac1{\omega_k}\ln
\frac{(\omega_k-\omega_{0k})^2+e^{\beta_0\omega_{0k}}(\omega_k+\omega_{0k})^2}{
(\omega_k+\omega_{0k})^2+e^{\beta_0\omega_{0k}}(\omega_k-\omega_{0k})^2}.}
The effective temperature obtained in this way turns out to be a function of
the momentum: 
in a free theory the different momentum modes do not interact with each
other and there is therefore no reason why they should all acquire the same
temperature. As before this property is not new; it is also true for a pure
quantum quench where 
\cite{2} \looseness=-1
\eq[cond1]{
\bef^*(k)=\frac2{\omega_k}\ln \frac{\omega_k+\omega_{0k}}{|\omega_k-\omega_{0k}|}.}
Note that the three temperatures $\beta_0, \bef$ and $\bef^*$
satisfy the following symmetric relation 
\eq[cond2]{\tanh\frac{\bpq(k) \omega_{k}}{2} \tanh \frac{\beta_0 \omega_{0k}}{2} = \tanh \frac{\beff(k) \omega_{k}}{2}}

Let us investigate the main features of the effective temperature. 
We can distinguish the following limiting cases:
\begin{itemize}
\item  {Cold  
quench $\omega_{0k}\gg \beta_0^{-1}$:
we asymptotically reproduce the pure quantum quench result (\ref{cond1}).} 
\item Hot quench $\beta_0^{-1} \gg \omega_{0k}$: 
\eq{\bef(k)=\frac{2\beta_0}{1+\omega_k^2/\omega_{0k}^2 }.}
For $\omega_{0k}>\omega_{k}$, this leads to the  {interesting}
conclusion that the system becomes `colder' after the quench,
unlike the pure quench case where the system becomes `hotter'
since the temperature rises from zero to a finite value. In
particular in the deep quench limit $\omega_{0k}\gg \omega_{k}$
the final temperature is half of the initial one.  {This can be
explained by the comments after (\ref{E}) along with the
fact that from (\ref{E0}) the energy of a harmonic oscillator at
thermal equilibrium with temperature $\beta^{-1}\gg \omega$ 
tends
to the classical value $\beta^{-1}$. }\looseness=-1
\item Critical evolution $m=0$: In this case, the zero-momentum effective
temperature has the simple form \eq[betacri] {\bef (k=0)
=\frac{4}{m_0} \tanh \frac{\beta_0 m_0}{2},} showing explicitly the
crossover between cold and hot quench. In particular, when $\beta_0 m_0= 3.83002\dots$ the effective temperature is equal to the initial one. 
\end{itemize}

\begin{figure}[tbp]
        \centering
                \includegraphics[width=\columnwidth]{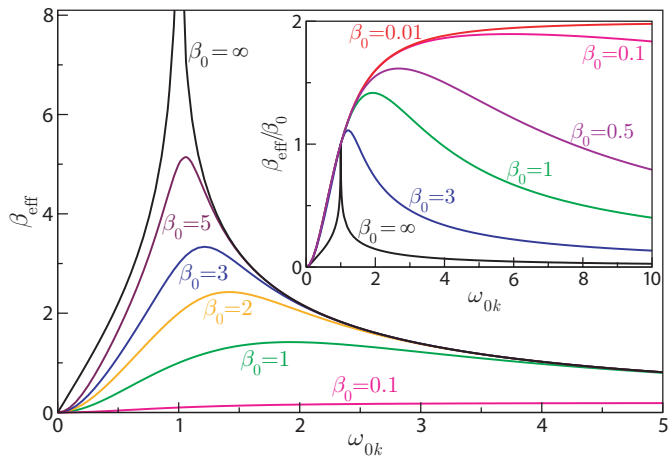} 
        \caption{\emph{ {Effective temperature after a thermal quantum quench as
            a function of $\omega_{0k}$ for several values of $\beta_0$ at
            $\omega_{k} = 1$. Inset: $\bef/\beta_0$ always for $\omega_{k} = 1$.}}} 
        \label{fig1}
\end{figure}

We point out that the relation $\bef=2\beta_0$
for the hot critical quench, is the same obtained (for any
observable) from the classical fluctuation-dissipation relation in
the ageing regime in a gaussian field theory \cite{15}.
Fig.~\ref{fig1} shows plots of $\bef$ as a function of the frequency  {$\omega_{0k}$} 
for several values of $\beta_{0}$. 
Note that the assumption $\omega_{0k} \gg \omega_k$ requires both
$m\ll m_{0}$ and $k \ll m_{0}$. Although the small momentum modes
indeed play the most important role in determining the stationary
behaviour of the system, one has to check the contribution of
large momenta as well and this is taken into account in the
following calculation.

\section{An alternative effective temperature}
Ideally the effective temperature should be a single number
describing the asymptotic state after the quench. It is then
important that this number should not depend on the particular
quantity we measure. The zero-momentum $\bef(k)$ is by definition
in a free theory a good measure for the large-distance behaviour
of any correlation function (we have derived it from the two-point
function, but in a free theory this is enough to calculate all of
them). Also in a conformal field theory the zero-momentum
effective temperature is well-defined and independently observable
\cite{2}. However, one could wonder whether physical
quantities involving higher momentum modes can spoil this result. 
To this end, we propose another estimate of the effective
temperature
by the comparison of the field fluctuations $\langle
\phi^2(x=0,{t\to\infty}) \rangle$ in the two cases: after the
quench and at thermal equilibrium
\ee[bar]{\int d^d k \left[ \frac{\omega_{0k}}{4} \left(\frac{1}{\omega_{0k}^2} +
\frac{1}{\omega_k^2}\right)  \coth \frac{\beta_0 \omega_{0k}}{2}  -\frac{1}{2 \omega_k} \right]  = \nn \\
 = \int d^d k \frac{1}{\omega_k (e^{\bar\beta \omega_k}-1)}. }
Such an estimate 
involves an average over all momentum modes.
For this reason we will call it `average' effective temperature and denote
it by $\bar\beta$. In the massless limit, both integrals are infrared
divergent in 1$d$ or 2$d$, but the two sides compensate each other.

Eq.~(\ref{bar}) can be solved numerically or even analytically in
the interesting limit $\beta_0^{-1} \gg m_0 \gg m$ mentioned
above. Then we expect $\bar\beta$ to be small too and we can use
the relevant asymptotic form of the thermal integral
\eq{I_{th} = \int_0^\infty{\frac{k^{d-1} dk}{\omega_k
(e^{\bar\beta \omega_k}-1)}},\nn} which is easily worked out for any
number of spatial dimensions $d$ (see table~\ref{tab1}). The
integral of the left hand side of (\ref{bar}) can be split into
two parts: the pure quench part which can be calculated exactly
 {\eq{I_{1} = \int_0^\infty{k^{d-1} dk \left[
\frac{\omega_{0k}}{4} \left(\frac{1}{\omega_{0k}^2} +
\frac{1}{\omega_{k}^2}\right) -\frac{1}{2 \omega_{k}} \right] },\nn}  } 
and the remaining $\beta_0$-dependent part which turns out to dominate in
the above limit 
\eq{I_{2}= \int_0^\infty{k^{d-1} dk
\frac{\omega_{0k}}{4} \left(\frac{1}{\omega_{0k}^2} +
\frac{1}{\omega_{k}^2}\right) \left[\coth \frac{\beta_0
\omega_{0k}}{2}  - 1\right] }. \nn} Table~\ref{tab1} summarizes the
asymptotic behaviour of the above integrals in the hot deep quench
limit. The comparison shows that in 1$d$ and 2$d$, $\bar\beta = 2
\beta_{0}$ as happens for $\bef(k)$ for small momenta. In 3$d$
however, $\bar\beta = \beta_{0}$ i.e. the average temperature
tends to the initial temperature instead of its half.
\begin{table}
\caption{\emph{Asymptotic behaviour of the integrals $I_{1},I_{2}$ and $I_{th}$ 
in the limit $\beta_0^{-1} \gg m_0 \gg m$.}}
\label{tab1}
\centering
\begin{tabular}{|l||c|c|c|}
\hline
$d$& 1 & 2 & 3 \\
\hline
$I_{1}$                         & ${\pi m_0}/{8 m}$                             & $-{m_0} \ln(m/m_0) /{4}$                                             & ${m_0^2}/{8}$                                                 \\
$I_{2}$                         & ${\pi}/{4 \beta_0 m}$                 & $-{\ln (\beta_0 m) }/{2 \beta_0}$    & ${\pi^2}/{6 \beta_0^2}$               \\
$I_{th}$        & ${\pi}/{2 \bar\beta m}$               & $-{\ln (\bar\beta m)}/{\bar\beta}$   & ${\pi^2}/{6 \bar\beta^2}$ \\
\hline
\end{tabular}
\end{table}
As a conclusion we observe that, even though the effective
temperature for small momenta is always half of the initial one,
the averaging of momenta involved in the calculation of the field
fluctuations and the fact that in $3d$ the dominant role of small
momenta is reduced, result in the average effective temperature
remaining the same as the initial one. Note also that in 2$d$
there are logarithmic corrections to the asymptotic behaviour
which could render comparison with data difficult.

Interestingly enough, in the above limit $\bar\beta=\bef(k=0)$ in 1$d$ and 2$d$ (and, as we will see, in any dimension for a lattice model for a sufficiently deep quench). This suggests that $\bef$ could have a sounding physical meaning even beyond the small momentum 
features. 

\section{Two-point correlation in real space}

It is worth to give a quick look at the equal-time correlation in real
space. The reason is twofold. Firstly we can explicitly see how the
zero-momentum effective temperature $\bef (k=0)$ describes its large distance
asymptotic. Second, we can understand how the pure quench horizon effect
\cite{4,1,2} is smoothed by finite temperature.

\begin{figure}[htbp]
        \centering
        \includegraphics[width= \columnwidth]{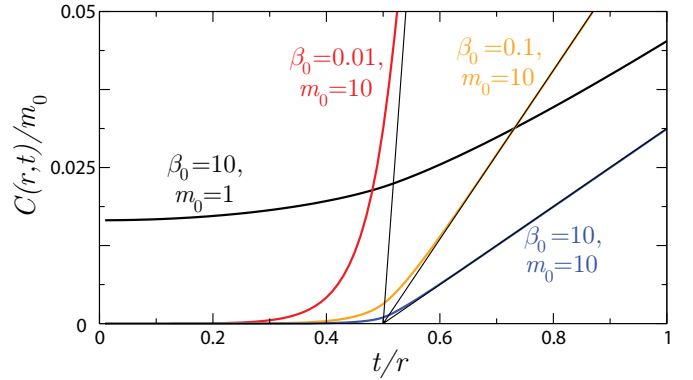}
        \caption{\emph{Smoothing of the horizon for real-space correlation.
        All curves refer to critical evolution $m=0$ and $d=1$.
         {The straight lines are the asymptotic behaviour.} }}
        \label{Fighor}
\end{figure}

 {From (\ref{pro}) the real space propagator for $t_1=t_2=t$ is 
\eq{
\int{ \frac{d^d k}{(2\pi)^d} e^{i \bf{ k \cdot r}}  \frac{\omega_{k}^{2} +  \omega_{0k}^{2} + (m^2-m_0^2) \cos 2  \omega_{k} t}{{4 \omega_{k}^{2}  \omega_{0k}}} \coth\frac{\beta_0 \omega_{0k}}{2}.\nn}} 
where, compared to the pure quench \cite{2}, there is only the additional
$\coth$ factor.} 
Let us firstly consider the massless (conformal) evolution with $m=0$ and $d=1$.
For large $r,t$ and $t>r/2$ a saddle point argument gives that the correlation function
is linear with a slope given by $\bef(0)$ as  {$C(r,t) \sim {\bef(0)}^{-1} (t-r/2) $}. 
Fig.~\ref{Fighor} shows how this limit is reached for several quenches.
Notice that in this regime the full correlation function
is described only by $\bef(0)$, that is a single number encoding all features
of the quench.
The massive evolution gives a less trivial saddle point that
superimposes slowly decaying large oscillations that are analogous to those
found in the pure quench \cite{2} and will not be discussed further.

From Fig.~\ref{Fighor} it is also evident how the finite temperature smooths
the horizon, which is not sharp as in the pure case. The interpretation
of this fact is straightforward: quasi-particles emitted at $t=0$ from a distance
smaller than the thermal correlation length $\propto \beta_0$
are entangled and generate correlations between two points at distance $r$
faster than particles emitted from the same point (or inside the true
correlation length $m_0^{-1}\ll \beta_0$). One is then tempted to conclude that
the effect of the temperature is similar to $m_0^{-1}$, but this is only
partially true. In fact a finite $m_0$, not only smooths the horizon, but
also produces a shift  in the zero time correlation (see the corresponding
curve in the figure).

\section{Free bosons on the lattice}

Let us now consider a system of free bosons on a lattice whose
 hamiltonian, if expressed in terms of the field operators, takes
the form \ee[ham2]{H = \sum_{k \in \text{BZ}}{\(\frac{1}{2}
\pi_{k}^{2} + \frac{1}{2} \omega^{2}_{k} \phi_{k}^{2}\)},} 
 {with a lattice dispersion relation, 
typically of the form}
\eq[latdisp]{\omega_{k}^{2} = m^2 c^{4} + 2 \frac{c^{2}}{a^{2}}
\sum\limits_{j}(1-\cos k_{j} a),}
where $m$ is again the energy gap,  $a$ is the lattice spacing and
$j$ enumerates the $d$ space coordinates. For convenience we set
$c=1$ and $a=1$. 

The only essential difference with the previous treatment is in
the dispersion relation and the fact that the momentum $k$ now
takes values in the first Brillouin zone  (BZ) $[-\pi,+\pi]^{d}$.
This obviously affects all quantities that involve a sum over all
momenta, but not the small momentum ones. First of all we have to
verify that the $(t_{1}+t_{2})$-dependent part of the propagator
is still decaying with time. Although there are now additional
stationary points in $\omega_{k}$ that determine the asymptotic
behaviour of the $k$-integral (specifically the edges of the
Brillouin zone) their contributions also
decay with time as before. Therefore thermalization also occurs in
the lattice. Secondly the computation of the average effective
temperature is easier since the natural cutoff introduced by the
lattice spacing simplifies the asymptotic analysis of the momentum
integrals. More explicitly, in the
deep quench limit, 
we can assume that $m_{0}$ is much larger than all $k$ in the
Brillouin zone. Then the left hand side of (\ref{bar}) can be
written in any dimension as ${ ({m_{0}}/{4}) \coth
({\beta_0 m_{0}}/{2})  \sum_{k}
({1}/{\omega_k^2})}$ and the right hand side becomes in the same
limit ${{\bar{\beta}}^{-1} \sum_{k}
({1}/{\omega_{k}^{2}})}$. Thus the average
effective temperature is 
\eq[cross]{ {\bar \beta}^{-1} =
\tfrac{1}{4} m_{0} \coth ({\beta_0 m_{0}}/{2}), } 
 {which is the same as (\ref{betacri}), showing that in the lattice the
two effective temperatures are equal in any dimension.}

\section{Other quenches}

Up to now we have considered only quenches of the mass parameter
 {of a relativistic dispersion relation.} 
Other
possibilities can also be studied in exactly the same way using
(\ref{pro}) as the starting point, checking the thermalization
condition and estimating the effective temperature in various
limits. For example one could quench the speed of sound $c$
or assume a classical dispersion relation 
${\omega_{k} = \Delta + {k^{2}}/{2 m^{*}}}$ 
and quench independently the
energy gap $\Delta$ or the effective mass $m^{*}$. However not all
of these possibilities are physically meaningful in the context of
a continuous field theory where the small scale degrees of freedom
have been completely ignored as presumably not playing a crucial
role. This means that the quench should not alter the large-$k$
behaviour of the spectrum. This explains the emergence of
ultraviolet divergences in the field fluctuations, in any
dimension, when one considers quenches of the speed of sound $c$. 
Similar divergences occur at a quench of the effective mass
$m^{*}$ in the classical dispersion relation above, except
for 1$d$ and $\Delta \neq 0$ where we find that the system
thermalizes.
This also happens for all dimensions in the case of a quench of
the energy gap $\Delta$ only. \looseness=-1

The conclusion is that as long as the quench does not alter the
large-$k$ behaviour of the spectrum in such a way as to cause
divergences, thermalization occurs for a broad class of gapped and
gapless dispersion relations. However the two simple types of
dispersion relations we considered so far are not sufficiently
adjustable to allow for any quench that would not affect the
large-$k$ behaviour, other than a quench of the energy gap.
Conversely, in lattice models, because of the presence of a natural
momentum cutoff preventing divergences, more elaborate quenches
are possible and well-defined. If for example we perform a (pure)
quench of the speed of sound of the lattice dispersion relation
${\omega_{k}^{2} = \Delta^2 + 2 {c^{2}} \sum_{j}(1-\cos k_{j}),}$
with $\Delta \neq 0$ kept fixed, then the propagator thermalizes
as before, but the momentum distribution of quasiparticles as
given by the stationary part of the propagator exhibits a maximum
at a nonzero $k$ and it vanishes at $k=0$. This means that after
such a quench the initial pure state feeds higher excitation
levels rather than the lowest ones which remain empty. 
The case $\Delta = 0, d=1$ is again special: the propagator does not tend to a
constant value, i.e. it does not thermalize in the usual sense,
but rather increases for large times as $\ln (t/a)$. However, as
mentioned also earlier, in the gapless regime the correlation
functions of vertex operators are physically more
meaningful\footnote{In 
  fact, in the Luttinger model \cite{9}, the logarithmic behaviour of the
  boson propagator leads to a power-law time decay of the fermionic two-point
  correlation function.}. \looseness=-1

More complex quenches are possible in free lattice models with
non-trivial spectra. In any case the procedure for the computation
of the time evolution is the same and general expressions can 
be readily obtained. Let us consider a system of free
bosons described by a general hamiltonian of the form \eq[h]{H =
\sum_{k}{A_k a_k^\dagger a_k + \tfrac{1}{2} B_k (a_{-k}^\dagger
a_k^\dagger + a_{-k} a_k) },} where $A_k,B_k$ are even functions
of $k$ so that $H$ is hermitian.  Suppose that the initial
temperature is $\beta_0$ when we quench the parameters of the
model from $A_0, B_0$ to $A,B$ and that we wish to find the time
evolution of the momentum distribution $n_k(t) = \la a_k^\dagger
a_k \ra$. To this end we need to diagonalise each of the two hamiltonians by means of a Bogoliubov transformation. Then to calculate $n_k(t) = \text{Tr}\{e^{-\beta_0 H_0}
e^{+iHt} a_k^\dagger a_k e^{-iHt}\}/Z(\beta_0)$ we first have to
express $a_k^\dagger a_k$ in the basis in which $H$ is diagonal 
so that the action of the time evolution
operator $\exp(-iHt)$ can be worked out easily. Next we have to
rewrite the operators in the basis in which $H_{0}$ is diagonal so that the
expectation value on the initial thermal state can also be worked
out, using the fact that the momentum distribution in this basis is equal to the
Bose-Einstein distribution $1/(\exp(\beta_0 E_{0k})-1)$. We finally obtain
\eq{n_k(t) = \frac{A_{0k}}{E_{0k}} \coth\frac{\beta_0 E_{0k}}{2}
\(\frac{B_k^2}{E_k^2} \sin^2 E_k t + \frac{1}{2} \) -
\frac{1}{2}.} As usual, the momentum distribution does not
equilibrate itself, but since any measurement is done on a finite
region of real space, the oscillating part of $n_{k}(t)$ is
unobservable for sufficiently large times due to the momentum
averaging that comes into play. This argument however relies on
the functional form of the excitation spectrum $E_k$ and
especially its stationary points. If this is such that the
oscillating part can be neglected, the observable distribution is
the stationary part \eq{\overline{n}_k = \frac{A_{0k} A^{2}_{k}}{2
E_{0k} E^{2}_{k}} \coth\frac{\beta_0 E_{0k}}{2} - \frac{1}{2}.} 

If we had a system of fermions instead of bosons then we could use
the same form (\ref{h}) for the hamiltonian but with $B_{k}$ being
an odd function of $k$ and imaginary. For clarity let us redefine
it by the substitution $B_k \to i B_k$ so that it is real. The energy
spectrum would then be $E_{k}^2 = A_{k}^2 + B_{k}^2$ and following
the same procedure we would find \eq{n_k(t) =
\frac{A_{0k}}{E_{0k}} \tanh\frac{\beta_0 E_{0k}}{2} \(
\frac{B_k^2}{E_k^2} \sin^2 E_k t - \frac{1}{2}\) + \frac{1}{2},}
and \eq{\overline{n}_k = \frac{1}{2} - \frac{A_{0k} A^{2}_{k}}{2
E_{0k} E^{2}_{k}} \tanh\frac{\beta_0 E_{0k}}{2}. }  {The above
expressions can be used to determine the momentum distribution
after any quench of a free lattice model.} 
Note that these results are in agreement with earlier studies (for example \cite{8,10}).

\section{Conclusions}

We have explored various cases of quenches in free models that
lead to stationary behaviour for large times and investigated its
effective thermal properties, also suggesting a way to estimate an
average effective temperature from the field fluctuations. We
showed that if the initial temperature of the system is nonzero
and higher than other parameters then the quench can lower it
significantly. 
 {We expect that several of our findings are not only valid for systems
admitting a free particle representation, but are general aspects
of thermal quantum quenches.} 

\acknowledgments
We thank Mihail Mintchev and Michael Bortz for fruitful discussions.
This work was supported in part by EPSRC grant EP/D050952/1. S. Sotiriadis
acknowledges financial support from St John's College, Oxford, and the
A. G. Leventis Foundation.


\end{document}